\documentclass[conference]{IEEEtran}
\IEEEoverridecommandlockouts

\usepackage{cite}
\usepackage{amsmath,amssymb,amsfonts}
\usepackage{algorithmic}
\usepackage{graphicx}
\usepackage[hidelinks]{hyperref}
\usepackage{textcomp}
\usepackage{xcolor}
\usepackage{algorithm}
\usepackage{subcaption}
\def\BibTeX{{\rm B\kern-.05em{\sc i\kern-.025em b}\kern-.08em
    T\kern-.1667em\lower.7ex\hbox{E}\kern-.125emX}}
\begin{document}

\title{
Admission Control with Reconfigurable Intelligent Surfaces 
for 6G Mobile Edge Computing
}


\author{
\IEEEauthorblockN{Ye Zhang\textsuperscript{1}, Baiyun Xiao\textsuperscript{1}, Jyoti Sahni\textsuperscript{1}, Alvin Valera\textsuperscript{1}, Wuyungerile Li\textsuperscript{2,*}, Winston K.G. Seah\textsuperscript{1}}
\IEEEauthorblockA{\textsuperscript{1}Victoria University of Wellington, Kelburn Parade, Wellington 6012, New Zealand\\
\textsuperscript{2}Inner Mongolia University, 235 West University Road, Hohhot 010021, China\\
\textsuperscript{*}Corresponding author: gerile@imu.edu.cn}
}

\maketitle

\begin{abstract}
As 6G networks must support diverse applications with heterogeneous quality-of-service requirements, efficient allocation of limited network resources becomes important. This paper addresses the critical challenge of user admission control in 6G networks enhanced by Reconfigurable Intelligent Surfaces (RIS) and Mobile Edge Computing (MEC). We propose an optimization framework that leverages RIS technology to enhance user admission based on spatial characteristics, priority levels, and resource constraints. Our approach first filters users based on angular alignment with RIS reflection directions, then constructs priority queues considering service requirements and arrival times, and finally performs user grouping to maximize RIS resource utilization. The proposed algorithm incorporates a utility function that balances Quality of Service (QoS) performance, RIS utilization, and MEC efficiency in admission decisions. Simulation results demonstrate that our approach significantly improves system performance with RIS-enhanced configurations. For high-priority eURLLC services, our method maintains over 90\% admission rates even at maximum load
, ensuring mission-critical applications receive guaranteed service quality.
\end{abstract}

\begin{IEEEkeywords}
6G, reconfigurable intelligent surfaces, mobile edge computing, admission control, resource allocation.
\end{IEEEkeywords}

\section{Introduction}
The rapid evolution of mobile networks towards 6G brings unprecedented opportunities for next-generation communication systems. Mobile Edge Computing (MEC) has emerged as a promising paradigm to bring computational resources closer to end-users, reducing latency and alleviating backhaul congestion. However, MEC alone cannot address all the challenges in 6G networks, particularly in non-line-of-sight (NLOS) scenarios with poor channel conditions. Reconfigurable Intelligent Surfaces (RIS) represent a revolutionary technology in wireless communications that can transform the propagation environment into a controllable domain~\cite{liu2021reconfigurable}. By passively reflecting incident electromagnetic waves in desired directions, RIS can significantly enhance signal quality, extend coverage, and improve overall network performance without requiring additional power-hungry active components. 
Despite the promising advantages, effective resource management in RIS-enhanced MEC systems remains a critical challenge. Users in 6G networks have heterogeneous requirements in terms of computational demands, communication quality, and latency constraints. Furthermore, RIS resources must be carefully allocated to maximize system utility~\cite{aboagye2022ris}.

Recent research has made significant progress in optimizing RIS-enhanced wireless networks. He et al.~\cite{he2023reconfigurable} proposed an efficient framework for user admission control in RIS-enhanced networks. 
Similarly, Li et al.~\cite{li2021joint} developed joint admission control and beamforming techniques to maximize the number of served users while minimizing power consumption. Building on these advances, Xiong et al.~\cite{xiong2025joint} introduced a sigmoid-based approximation approach for admission control in RIS networks. 
Zhang et al.~\cite{zhang2022user} investigated the integration of RIS with user scheduling in multiuser Multiple-Input Single-Output (MISO) systems. 
Additionally, Zivuku et al.~\cite{zivuku2023joint} explored multi-slot scheduling combined with RIS-enhanced precoding to maximize user admission in smart city scenarios. The work by Fu et al.~\cite{fu2019intelligent} further demonstrated how intelligent reflecting surfaces can enhance spatial multiplexing capabilities. Furthermore, Wu et al.~\cite{wu2019intelligent} have shown that joint active and passive beamforming optimization can dramatically improve signal quality in RIS-enhanced networks.

However, several critical challenges remain unaddressed in the existing literature. First, most current approaches fail to consider the angular characteristics of RIS in the admission control process. Second, existing works typically treat users as homogeneous entities without accounting for diverse service priorities. 
Third, the spatial grouping of users for efficient RIS resource allocation has not been sufficiently explored. 
Finally, there is a lack of comprehensive frameworks that jointly consider RIS capabilities, MEC resources, and user priorities in admission decisions. 
The main contributions of this paper are as follows: 
\begin{itemize}
    \item We formulate the problem of user admission control in RIS-enhanced MEC environments, considering the spatial characteristics of RIS-enhanced communication, heterogeneous user priorities, and resource constraints.  
    \item We propose a user grouping strategy that efficiently allocates RIS resources by clustering users with similar angular requirements. 
    \item We develop a framework that maximizes system utility by considering user communication quality, computational requirements, and RIS units in admission decisions.
\end{itemize}

The rest of this paper is organized as follows: Section~\ref{sec:System} presents the system model. Section~\ref{sec:Algorithm} details the problem formulation and our proposed algorithm. Section~\ref{sec:Validation} provides simulation results to validate the effectiveness of our approach. Finally, Section~\ref{sec:Conclusions} concludes the paper and discusses future research directions.

\section{System Model}\label{sec:System}

In this section, we present a comprehensive system model for RIS-enhanced mobile edge computing in 6G networks. 

\subsection{Network Architecture}

We consider a 6G wireless network comprising a base station (BS) equipped with MEC capabilities, RIS, and a set of users $U = \{u_1, u_2, \ldots, u_N\}$ generating dynamic service requests. The RIS panels consist of $M$ passive reflecting elements that can be configured to enhance signal propagation between the BS and users.
The network consists of the following components: (1) Base Station: Located at coordinates $(x_B, y_B, z_B)$ with computational capacity $C_{max}$ and communication resources $B_{max}$. (2) RIS Panels: Located at coordinates $(x_R, y_R, z_R)$ with $M$ configurable reflecting elements. (3) Users Requests: Each user $u_i$ is located at $(x_u^i, y_u^i, z_u^i)$ and characterized by priority level $P_i$, arrival time $t_i$, computational requirements $C_i$, bandwidth requirements $B_i$, and maximum tolerable delay $T_{max,i}$. The user requests are categorized into three service types with distinct QoS requirements: Enhanced Ultra-Reliable Low-Latency Communication (eURLLC), Further Enhanced Mobile Broadband (feMBB), Ultra Massive Machine Type Communication (umMTC)~\cite{wang2023road}. 

\subsection{RIS Angular Characteristics}

The effectiveness of RIS-enhanced transmission depends significantly on the angular alignment between users, RIS, and the BS. The reference direction is defined by the vector from RIS to BS, given by
\begin{equation}
    \mathbf{V}_{RB} = (x_B - x_R, y_B - y_R, z_B - z_R).
\end{equation}
For each user $u_i$, the direction vector from RIS is
\begin{equation}
    \mathbf{V}_{Ru}^i = (x_u^i - x_R, y_u^i - y_R, z_u^i - z_R).
\end{equation}
The angular deviation between user $u_i$ and the reference direction is given by
\begin{equation}
    \cos\theta_i = \frac{\mathbf{V}_{RB} \cdot \mathbf{V}_{Ru}^i}{\|\mathbf{V}_{RB}\| \|\mathbf{V}_{Ru}^i\|}.
\end{equation}
This allows us to obtain $\theta_i$ using
\begin{equation}
    \theta_i = \arccos(\cos\theta_i).
\end{equation}

This angular information is crucial for determining which users can effectively benefit from RIS enhancement~\cite{tewes2023comprehensive}. Users whose angular deviation exceeds a certain threshold will experience limited signal improvement, making them less suitable candidates for RIS-enhanced transmission.

\subsection{Computational Model}
For tasks offloaded to the MEC server, each user $u_i$ has associated computational requirements $C_i$ and data size $D_i$. The total delay $T_i$ experienced by user $u_i$ comprises both processing and transmission components:
\begin{equation}
    T_i = T_{proc,i} + T_{trans,i}
\end{equation}

Processing delay $T_{proc,i}$ depends on the computational resources $f_i$ allocated to user $u_i$, while transmission delay $T_{trans,i}$ is determined by the data size and achievable data rate $R_i$. Resource allocation must satisfy the MEC server's capacity constraint:
\begin{equation}
    \sum_{i=1}^{N} x_i \cdot f_i \leq F_{total}
\end{equation}
where $x_i$ is the binary admission decision variable and $F_{total}$ represents the total computational capacity of the MEC server.

\section{Admission Control Algorithm}\label{sec:Algorithm}

In this section, we propose an efficient algorithm for user admission and RIS resource allocation that maximizes system utility while ensuring QoS requirements are satisfied.

\subsection{User Filtering and Candidate Selection}

The first phase filters users based on their angular deviation from the RIS reference direction, as RIS enhancement is most effective when users are within a certain angular range relative to the RIS-BS direction.
    \subsubsection{Angular Deviation Computation} For each user $u_i$, calculate the angular deviation $\theta_i$ from the reference direction using Equations (3) and (4).
   
    \subsubsection{Candidate Set Formation} Select users whose angular deviation is within an acceptable range
    \begin{equation}
        U' = \{ u_i \in U \ | \ |\theta_i| \leq \theta_{\text{range}} \}
    \end{equation}
    where $\theta_{\text{range}}$ is the maximum allowable angular offset.
The filtering process enables the algorithm to focus computational resources on users who can meaningfully benefit from RIS enhancement. Users outside this angular range will rely solely on direct transmission paths and are processed separately by the conventional admission control mechanism of the BS.

\subsection{Priority-Based User Grouping}
The second phase organizes filtered users into groups sharing similar angular characteristics to enable efficient RIS reflection while minimizing interference.

\subsubsection{Priority Queue Generation} 
Candidate users are sorted based on priority level $P_i$ and arrival time $t_i$:
\begin{equation}
    Q = \text{sort}(U', \text{key} = (-P_i, t_i))
\end{equation}
This ensures high-priority users are considered first, with earlier arrivals taking precedence among equal-priority users. Priority level $P_i$ is determined by service type (eURLLC, feMBB, or umMTC).

 \subsubsection{Group Formation}
     User grouping is performed based on angular similarity and resource constraints. First, we select the first user in the priority queue as the group leader $u_{\text{leader}}$ and set the base angle for the group as $\theta_{\text{base}}^i = \theta_{\text{leader}}$. Users are included in the group if their angular deviation satisfies $|\theta_i - \theta_{\text{base}}^i| \leq \theta_{\text{max}}$, where $\theta_{\text{max}}$ is the maximum allowed deviation within a group. This parameter controls the angular coherence within a group and affects the trade-off between group size and interference.

\subsubsection{RIS Resource Allocation} 
RIS resources are distributed proportionally to group priorities:
\begin{equation}
    r_i = \frac{P_G^i}{\sum_{j \in G} P_G^j} \cdot M
\end{equation}
where $P_G^i = \sum_{u_j \in G_i} P_j$ is the total priority of group $G_i$, ensuring groups with higher collective priority receive larger shares of RIS resources.

\subsection{Utility Maximization Framework}

The final phase determines the optimal admission decision and resource allocation by maximizing system utility.

   \subsubsection{User Utility Function} Define the utility for user $i$ as
    \begin{equation}
        U_i = \alpha_1 \cdot U_{\text{rate}, i} + \alpha_2 \cdot U_{\text{delay}, i}
    \end{equation}
    where $\alpha_1$ and $\alpha_2$ are weighting factors that prioritize rate or delay performance based on service requirements. 
    The component utilities are defined as
    \begin{equation}
        U_{\text{rate}, i} = \frac{R_i}{\max(R_i)}
    \end{equation}
    \begin{equation}
        U_{\text{delay}, i} = 1 - \frac{T_i}{T_{\text{max}, i}}.
    \end{equation}
    These normalized utility functions ensure that both rate and delay aspects contribute proportionally to the overall utility, regardless of their absolute values.

    \subsubsection{RIS Resource Utility} Quantify the benefit of RIS resource allocation using
    \begin{equation}
        R_{\text{RIS}} = \sum_{g=1}^{G} P_g \sum_{i \in g} z_i \cdot r_i
    \end{equation}
    where $z_i$ is a binary variable indicating RIS allocation to user $u_i$ (1 if allocated, 0 otherwise), $r_i$ is the number of RIS elements allocated, and $P_g$ is the priority weight of group $g$. This term encourages efficient allocation of RIS resources to high-priority groups.

    \subsubsection{Admission Penalty} Define a penalty for rejecting users to ensure fairness using
    \begin{equation}
        \text{Penalty} = \frac{\sum_{i=1}^{N} (1 - x_i) \cdot P_i}{P_{\text{max}}}
    \end{equation}
    where $x_i$ is the admission decision variable (1 if admitted, 0 if rejected) and $P_{\text{max}}$ is the maximum total priority value. This penalty increases when high-priority users are rejected, encouraging the algorithm to admit users with higher priority when resources are limited.

    \subsubsection{Optimization Problem} Formulate the overall problem as:
    \begin{equation}
        \!\!\max_{\{x_i, z_i, r_i\}} \sum_{i=1}^{N} \left( \gamma_1 \cdot P_i \cdot x_i \cdot U_i + \gamma_2 \cdot R_{\text{RIS}} - \gamma_3 \cdot \text{Penalty} \right)
    \end{equation}
    


\begin{algorithm}
\caption{RIS-Enhanced User Admission Control}
\small
\begin{algorithmic}[1]
\REQUIRE User set $U$, RIS parameters, system constraints
\ENSURE Admission decisions $x_i$, RIS allocations $z_i$, $r_i$

\STATE Calculate reference direction $\mathbf{V}_{RB}$
\FOR{each user $u_i \in U$}
    \STATE Calculate $\theta_i$
    \IF{$|\theta_i| \leq \theta_{\text{range}}$}
        \STATE Add $u_i$ to candidate set $U'$
    \ENDIF
\ENDFOR

\STATE Sort $U'$ by priority and arrival time to form queue $Q$
\STATE Initialize empty group set $G$
\WHILE{$Q$ is not empty}
    \STATE Select first user $u_{\text{leader}}$ from $Q$ as group leader
    \STATE Initialize new group $G_k = \{u_{\text{leader}}\}$
    \STATE Remove $u_{\text{leader}}$ from $Q$
    \FOR{each remaining user $u_j$ in $Q$}
        \IF{$|\theta_j - \theta_{\text{leader}}| \leq \theta_{\text{max}}$}
            \IF{adding $u_j$ to $G_k$ satisfies resource constraints}
                \STATE Add $u_j$ to $G_k$
                \STATE Remove $u_j$ from $Q$
            \ENDIF
        \ENDIF
    \ENDFOR
    \STATE Add $G_k$ to $G$
\ENDWHILE

\FOR{each group $G_k$ in $G$}
    \STATE Calculate group priority $P_G^k$
    \STATE Allocate RIS resources proportionally to $P_G^k$
\ENDFOR
\STATE Initialize $x_i = 0, z_i = 0, \forall i \in U$
\STATE Initialize available resources $C_{avail} = C_{max}, R_{avail} = M$
\REPEAT
    \STATE Find user $u_i$ with maximum marginal utility contribution
    \IF{admitting $u_i$ satisfies all constraints}
        \STATE Set $x_i = 1$
        \IF{$u_i \in U'$}
            \STATE Set $z_i = 1$
            \STATE Update $R_{avail} = R_{avail} - r_i$
        \ENDIF
        \STATE Update $C_{avail} = C_{avail} - C_i$
    \ENDIF
\UNTIL{no more users can be admitted}
\RETURN admission decisions $x_i$, RIS allocations $z_i$, $r_i$
\end{algorithmic}
\end{algorithm}

        The objective function maximizes the weighted sum of user utilities, RIS resource utility, and admission fairness. The weighting factors $\gamma_1$, $\gamma_2$, and $\gamma_3$ control the relative importance of these components and can be adjusted to achieve different performance objectives. The proposed algorithm is summarized in Algorithm 1. This algorithm provides a practical solution to the complex problem of user admission control in RIS-enhanced MEC environments. By considering both angular constraints and user priorities, it efficiently allocates resources to maximize system utility while ensuring QoS guarantees for admitted users.




\begin{figure*}[h]
  \centering
  \includegraphics[width=0.95\linewidth]{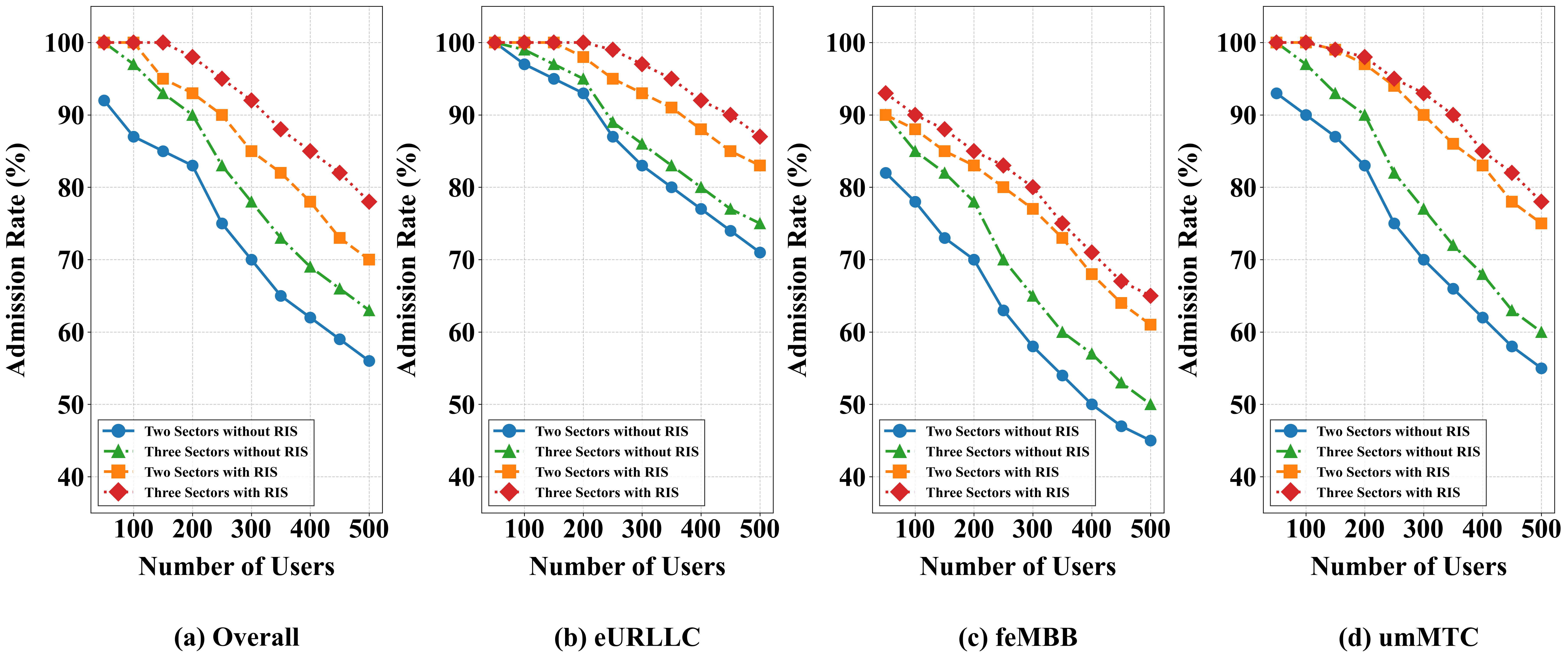}
  \caption{Admission rate comparison}
  \label{fig:1}
\end{figure*}
\section{Validation and Performance}\label{sec:Validation}
In this section, we present extensive simulation results to evaluate the performance of our proposed RIS-enhanced admission control algorithm. 

\subsection{Simulation Setup}
The base station is equipped with MEC capabilities and assumed to have sufficient resources to meet all user demands. Our focus is on the allocation of RIS resources to complement the communication demands where users are distributed across different sectors with varying densities leading to localized congestion. 
%
%
Users are categorized into three service types: eURLLC (highest priority), feMBB (medium priority), and umMTC (lowest priority). We assume the user density ratio for eURLLC, feMBB and umMTC of 2:6:1~\cite{khan2022urllc}. We randomly generated user requests with varying computational requirements, data sizes, and latency constraints according to the specified ranges for each service type. To simulate realistic network conditions, we deliberately created an unbalanced load distribution by generating more users in the congested sector (Sector 0) compared to other sectors, with a congestion ratio of 3.0, where in a two-sector system Sector 0 has 3 times more users than the other sector, or in a three-sector system Sector 0 has 1.5 times more users than all other sectors combined. We evaluated the performance of our proposed algorithm under four different configurations: two and three sectors with and without RIS. 

\subsection{Admission Rate Performance}

Fig.~\ref{fig:1} shows the admission rates for different network configurations across varying user densities. The overall admission rate demonstrates that our RIS-enhanced approach consistently outperforms traditional admission control methods without RIS. At maximum load (500 users), the three-sector configuration with RIS achieves an admission rate of approximately 78\%, which is significantly higher than the 55\% achieved by the two-sector configuration without RIS. The service-specific admission rates reveal that high-priority services (eURLLC) maintain higher admission rates even under heavy load conditions, which aligns with our design objective of prioritizing critical services. The three-sector configuration with RIS maintains over 85\% admission rate for eURLLC services even with 500 users, demonstrating the effectiveness of our approach in protecting high-priority traffic. Notably, the performance gap between RIS and non-RIS configurations widens as the number of users increases, highlighting the scalability advantages of our approach. This is particularly evident for umMTC services, where the admission rate difference reaches up to around 20 percentage points at high user densities.

\subsection{Sector-Specific Performance Analysis}

\begin{figure}[h] 
  \centering
  \includegraphics[width=\linewidth]{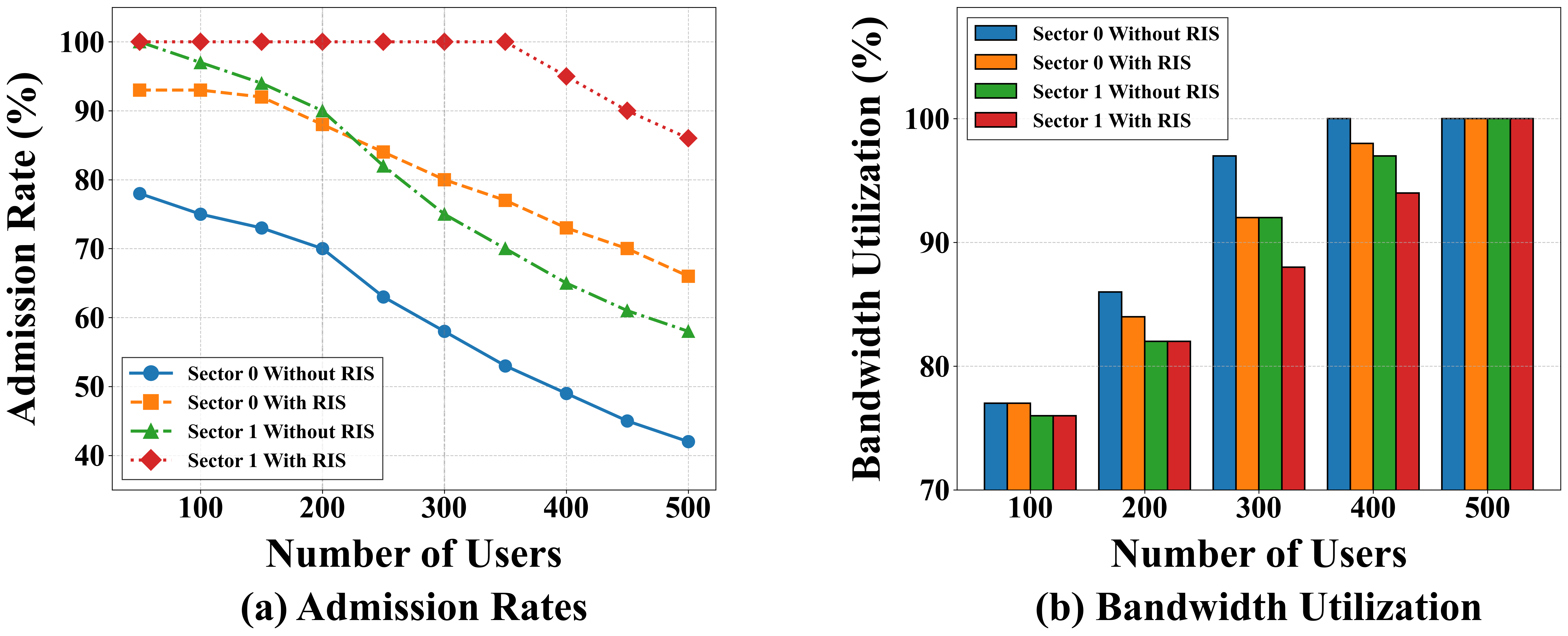}
  \caption{Two sectors comparison}
  \label{fig:2}
\end{figure}

\begin{figure}[h] 
  \centering
  \includegraphics[width=\linewidth]{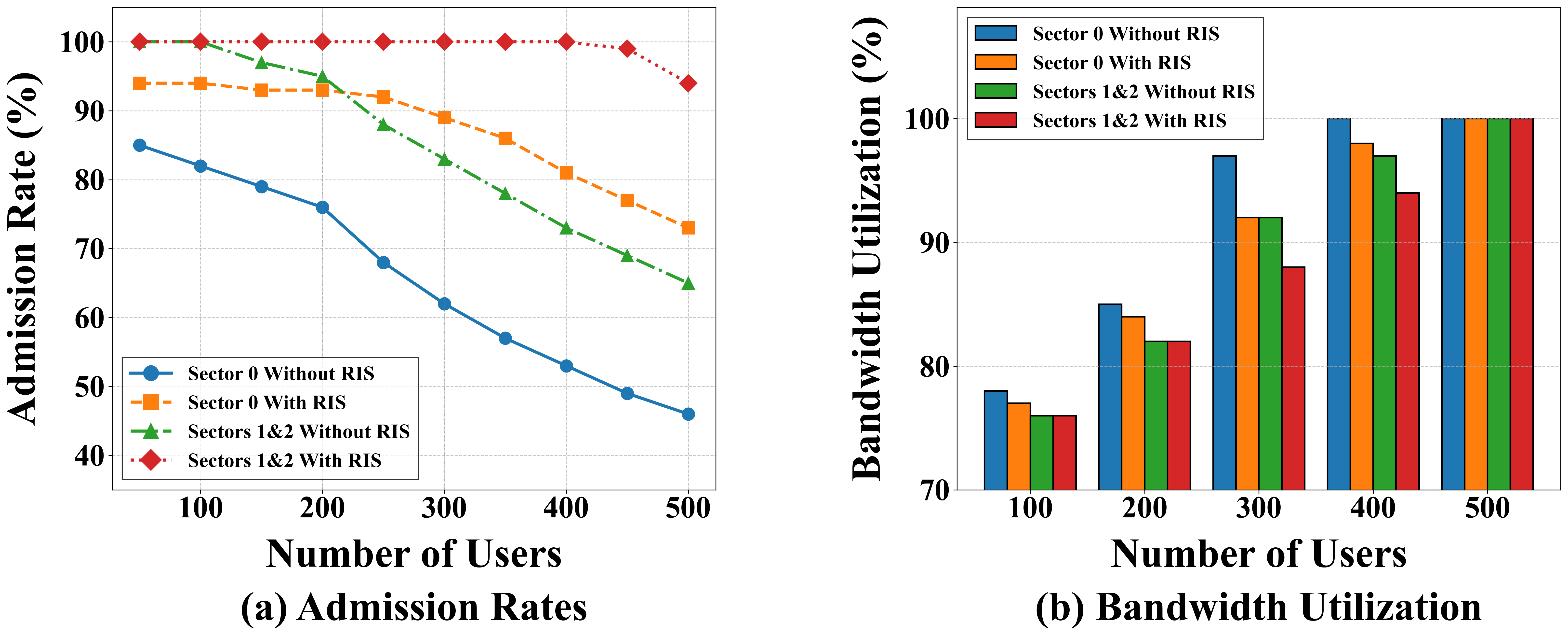}
  \caption{Three sectors comparison}
  \label{fig:3}
\end{figure}

Fig.~\ref{fig:2} and Fig.~\ref{fig:3} illustrate the performance of individual sectors in terms of admission rates and bandwidth utilization. 

In the two-sector configuration (Fig.~\ref{fig:2}), Sector 1 (the uncongested sector) shows a substantial improvement in admission rate with RIS assistance, increasing from 59\% to 86\% at maximum load. For the three-sector configuration (Fig.~\ref{fig:3}), the most significant improvement is observed in Sector 1 and 2, where RIS deployment increases the admission rate from 66\% to 94\% at maximum load. The congested sectors (Sector 0) maintain nearly identical performance, which aligns with our model's expectations. The bandwidth utilization plots in the right panels of Fig.~\ref{fig:2} and Fig.~\ref{fig:3} reveal that RIS-enhanced configurations achieve higher admission rates without proportionally increasing bandwidth consumption. This efficiency is particularly evident in the congested Sector 0, where our algorithm optimally distributes RIS resources to maximize signal enhancement while minimizing interference. Notably, the bandwidth utilization remains more efficient at user counts below 400 in the three-sector RIS configuration, demonstrating the resource-optimization capabilities of our approach.

\subsection{Latency Performance}

\begin{figure}[h] 
  \centering
  \includegraphics[width=0.8\linewidth]{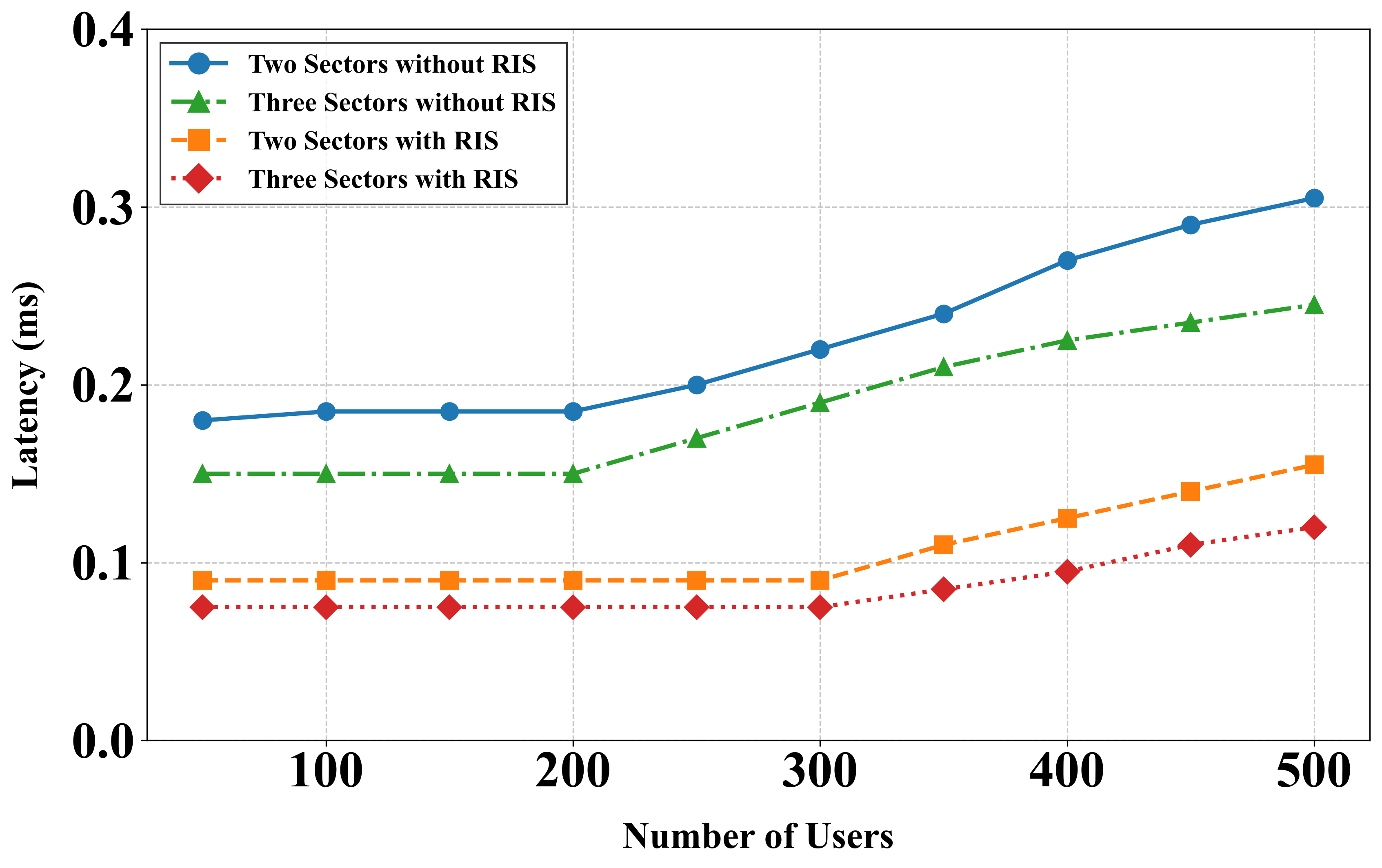}
  \caption{Latency comparison}
  \label{fig:4}
\end{figure}

Fig.~\ref{fig:4} presents the latency performance across different service types and network configurations. The overall latency shows that RIS-enhanced configurations achieve significantly lower latency compared to non-RIS configurations. At maximum load, the three-sector configuration with RIS achieves approximately 0.12ms overall latency, which is 60\% lower than the 0.3ms latency experienced in the two-sector configuration without RIS. This demonstrates that our approach successfully maintains latency within the QoS constraints defined in the optimization framework. 
This inflection point occurs much earlier (around 200 users) for configurations without RIS, demonstrating the capacity-enhancing capabilities of our RIS-enhanced approach and validating our utility-based optimization framework's effectiveness in resource allocation under varying load conditions.

The efficacy of our RIS-enhanced admission control algorithm fundamentally depends on the angular relationship between users, RIS and BS. Furthermore, our algorithm's user grouping mechanism successfully clusters users with similar angular requirements. 
Overall, these results demonstrate that our proposed RIS-enhanced admission control algorithm successfully leverages the spatial characteristics of RIS to improve network performance across all evaluated metrics.

\section{Conclusions}\label{sec:Conclusions}
In this paper, we addressed the challenge of user admission control in 6G mobile edge computing scenarios with RIS enhancements. We proposed an optimization framework that jointly considers user spatial characteristics, service priorities, and resource constraints to maximize the number of users served while maintaining QoS guarantees. Our simulation results validate the effectiveness of the utility-maximization approach, demonstrating significant improvements in admission rates and latency performance across different service types. The RIS-enhanced configurations consistently outperformed traditional approaches, particularly in congested scenarios, confirming that our utility-based resource allocation strategy successfully balances communication quality, computational efficiency, and RIS constraints in admission decisions. In our future work, we plan to investigate the benefits of multiple distributed RIS units compared to a single large RIS, and developing efficient coordination mechanisms for multi-RIS scenarios.

\section*{Acknowledgment}

The research by Ye Zhang and Baiyun Xiao were supported by the China Scholarship Council (Project ID: 202409150001 and 202308350020).

\bibliography{manuscript}
\bibliographystyle{IEEEtran}

\end{document}